# Archaeoastronomy in northern Chile:
# Andean churches of the Arica and Parinacota region

A. Gangui (1, 2), A. Guillén (3) and M. Pereira (3)

(1) Universidad de Buenos Aires, Facultad de Ciencias Exactas y Naturales, Argentina
(2) CONICET - Universidad de Buenos Aires, Instituto de Astronomía y Física del Espacio (IAFE), Argentina
(3) Fundación Altiplano, Chile

Abstract: We present the results obtained from the analysis of the precise spatial orientations of almost forty old colonial Christian churches located in the Arica and Parinacota region. This is an extended and difficult to travel area that received little attention from parish priests, and where one might expect some dialogue to have taken place between the Western tradition and the local Aymara culture in regard to the design and construction of temples. We also briefly comment on our plans to explore neighboring areas to the region here studied, which include many additional historical constructions that share similar cultural traditions.

**Introduction**

The spatial orientation of ancient Christian churches is one of the most characteristic features of their architecture (Dietz 2005). In Europe and in many remote sites where the missionaries arrived, there is a clear tendency to orient the altars of the temples in the solar range; namely, the axis of the temple, from the front door towards the altar, is aligned to those points on the horizon where the Sun rises on different days of the year. Among these days, there is a marked preference for those corresponding to astronomical equinoxes, when the axes point towards geographical east, the direction which down through the ages has symbolized the eschaton: the second coming of Christ in kingly glory (McCluskey 2015). However, even within the solar range, alignments in the opposite direction –with the altar to the west– are not uncommon, although they are exceptional because they do not follow the canonical pattern (see Gangui et al. 2016a, and references therein).

Soon after the Spanish settlement in the Viceroyalty of Peru, Francisco de Toledo decided to reorganize the territory and its population. He focused on the trade routes of the southern Andean area, the main objective of which was to organize the transport of silver from Potosi to the Pacific and also that of quicksilver from Huancavelica to the high Andean mines. On the way to the city of Arica, official maritime port for the merchandise since 1574, there appeared small villages and *tambos* with stable populations. Modest chapels and churches in this region emerged in strategic locations along the route which *trajinantes* marched to transport the precious metals from Potosi to Arica beaches, particularly around the Lluta and Azapa valleys.

**Churches of Arica and Parinacota – a description**

Religious architecture that began to materialize in the southern Andean region was characterized by elongated simple structures, rounded apses, stone walls without edging, pitched and thatched roofs (Fundación Altiplano 2012). In areas outside the temples there appeared catechetical crosses, arcaded atria, bell towers and small Miserere chapels. In the surrounding environments to the churches, the indigenous demands embodied in the sacred use of open spaces was clear, and formalized in a sort of cohabitation between indigenous and European morphological ideologies.

In Table 1 we show the data obtained in our campaign of fieldwork. The identification of the churches and their coordinates are presented along with their orientation (archaeoastronomical data): the measured azimuth (rounded to 1/2° approximation) and the angular height of the point of the horizon towards which the apse of the church is facing, as well as the corresponding computed astronomical declination. Regarding the construction dates provided for the churches, there is some ambiguity, as the documentary sources are scarce and often refer to ancient churches in the area that are no longer standing.

| Location | Name / Date | L (°/') South | l (°/') West | a (°) | h (°) | δ (°) | Patron saint date / orientation |
|---|---|---|---|---|---|---|---|
| (1) Azapa | San Miguel (1618; end of s. XIX) | 18/31 | 70/11 | 36½ | 15 | 42 | 29 Sep / ---- |
| (2) Poconchile | San Jerónimo (1618; end of s. XIX) | 18/27 | 70/04 | 156 | 15 | -66½ | 30 Sep / ----- |
| (3) Chitita | Virgen del Carmen (end of s. XIX) | 18/50 | 69/41 | 299½ | 11 | 24½ | 16 Jul / ----- |
| (4) Guañacagua | San Pedro (end of s. XIX) | 18/49 | 69/43 | 78½ | 14 | 6½ | 29 Jun / 7 Apr – 6 Sep |
| (5) Sucuna | San Antonio de Padua (end of s. XIX) | 18/51 | 69/27 | 275½ | 0 | 5½ | 13 Jun / 4 Apr – 10 Sep |
| (6) Saguara | La Santa Cruz (last third of s. XIX) | 18/54 | 69/30 | 266½ | 1 | -3½ | 3 May / 12 Mar – 2 Oct |
| (7) Pachica | San José (1618; beginning s. XVIII) | 18/55 | 69/37 | 76 | 9 | 10½ | 19 Mar / 18 Apr – 26 Aug |
| (8) Esquiña | San Pedro (1618; beginning s. XVIII) | 18/56 | 69/32 | 108½ | 15 | -21½ | 29 Jun / 14 Jan – 30 Nov |
| (9) Aico | San Antonio de Padua (end of s. XIX) | 18/48 | 69/28 | 69½ | 6 | 18 | 13 Jun / 12 May – 2 Aug |
| (10) Parcohaylla | San José (end of s. XIX) | 18/53 | 69/13 | 271 | 0 B | 1½ | 19 Mar / 24 Mar – 20 Sep |
| (11) Mulluri | Virgen de la Natividad (end of s. XIX) | 19/01 | 69/10 | 244 | 5 | -26 | 8 Sep / ----- |
| (12) Codpa | San Martín de Tours (1618; beginning s. XVIII) | 18/50 | 69/45 | 70½ | 13 | 14½ | 11 Nov / 29Apr – 15 Aug |
| (13) Timar | San Juan Bautista (1618; beginning s. XVIII) | 18/45 | 69/41 | 53½ | 8 | 31½ | 24 Jun / ----- |
| (14) Cobija | San Isidro Labrador (second half s. XIX) | 18/44 | 69/35 | 255½ | 5 | -15 | 15 May / 8 Feb – 4 Nov |
| (15) Timalchaca | Virgen de los Remedios (end of s. XIX) | 18/41 | 69/25 | 309½ | -1 | 38 | 21 Nov / ----- |
| (16) Tignamar | Virgen Asunción (1618; end of s. XIX) | 18/35 | 69/29 | 101½ | 6 | -12½ | 15 Aug / 17 Feb – 26 Oct |
| (17) Chapiquiña | San José Obrero (mid s. XX) | 18/24 | 69/32 | 63½ | 12 | 21 | 1 May / 27 May – 18 Jul |
| (18) Pachama | San Andrés Apóstol (s. XVIII) | 18/26 | 69/32 | 19½ | 10 | 57 | 30 Nov / ----- |
| (19) Saxamar | Santa Rosa de Lima (1618; s. XX) | 18/33 | 69/30 | 320½ | 3½ | 46 | 30 Aug / ----- |
| (20) Belén | Santiago Apóstol (1618; beginning s. XVIII) | 18/28 | 69/31 | 188½ | 16 | -81½ | 25 Jul / ----- |
| (21) Belén | Virgen de la Candelaria (second half s. XVIII) | 18/28 | 69/31 | 162½ | 19 | -73 | 2 Feb / ----- |
| (22) Socoroma | San Francisco de Asís (1618; end of s. XIX) | 18/16 | 69/36 | 15½ | 4½ | 63½ | 4 Oct / ----- |
| (23) Putre | Virgen Asunción (1618; end of s. XIX) | 18/12 | 69/34 | 352½ | 10 | 62 | 15 Aug / ----- |
| (24) Tacora | Virgen del Carmen (s. XVIII) | 17/46 | 69/43 | 289½ | 4 | 17½ | 16 Jul / 10 May – 4 Aug |
| (25) Airo | Santiago Apóstol (beginning s. XX) | 17/43 | 69/39 | 257½ | 3 | -12½ | 25 Jul / 16 Feb – 27 Oct |
| (26) Chapoco | San Martín de Tours (mid s. XX) | 17/43 | 69/34 | 203 | 6 | -65 | 11 Nov / ----- |
| (27) Putani | Virgen Inmaculada Concepción (s. XX) | 17/44 | 69/32 | 207 | 3½ | -60 | 8 Dec / ----- |
| (28) Pucoyo | Virgen del Rosario (beginning s. XX) | 17/46 | 69/25 | 253 | 6½ | -18 | 3 Oct / 30 Jan – 13 Nov |
| (29) Cosapilla | Virgen del Rosario (s. XVIII) | 17/46 | 69/25 | 219 | 13 | -52 | 3 Oct / ----- |
| (30) Guacollo | Santa Rosa de Lima (beginning s. XX) | 17/46 | 69/21 | 234 | 4 | -35½ | 30 Aug / ----- |
| (31) Parinacota | Virgen de la Natividad (s. XVIII) | 18/12 | 69/16 | 37 | 2½ | 48½ | 8 Sep / ----- |
| (32) Ungallire | Virgen del Carmen (end of s. XIX) | 18/12 | 69/16 | 249 | 3 | -20½ | 16 Jul / 18 Jan – 25 Nov |
| (33) Caquena | Santa Rosa de Lima (end of s. XIX) | 18/03 | 69/12 | 268 | 8 | -4 | 30 Aug / 10 Mar – 4 Oct |
| (34) Chucuyo | La Santa Cruz (end of s. XIX) | 18/13 | 69/18 | 270 | 0 B | 0 | ?? May / 22 Mar – 23 Sep |
| (35) Choquelimpie | Virgen Asunción (1618; end of s. XIX) | 18/19 | 69/16 | 197 | 20 | -74 | 15 Aug / ----- |
| (36) Ancuta | Virgen Inmaculada Concepción (end of s. XIX) | 18/27 | 69/12 | 252 | 1 | -17 | 8 Dec / 1 Feb – 11 Nov |
| (37) Guallatire | Virgen Inmaculada Concepción (1873) | 18/30 | 69/09 | 218 | 1 | -49 | 8 Dec / ----- |
| (38) Churiguaya | Virgen de la Candelaria (end of s. XIX) | 18/21 | 69/11 | 232 | 2 | -36½ | 2 Feb / ----- |

Table 1: Orientations for the heritage Andean chapels and churches of Arica and Parinacota. For each building, we show the location, identification (name, date of first reported reference of the building, when available, and most likely date of construction of the existing church), the geographical latitude and longitude (L and l), the astronomical azimuth (a) taken along the axis of the building towards the altar (rounded to 1/2° approximation), the horizon angular height (h) in that direction (0 B means horizon is blocked; we take h = 0°) and the corresponding resultant declination (δ).

In Fig. 1 we present the orientation diagram for the 38 measured churches and chapels. Among these, 6 are oriented in the northern quadrant; there are 8 oriented in the eastern quadrant (6 of them in the solar range) and 16 oriented in the western quadrant (11 of them in the solar range, with azimuths between 245.0° and 294.6°). Finally, there are 8 churches in the southern quadrant (azimuths between 135° and 225°). Based on their locations in the territory and on their number we think that our sample is representative of most of the temples in the region. In our data, a prominent orientation towards the western quadrant is apparent.

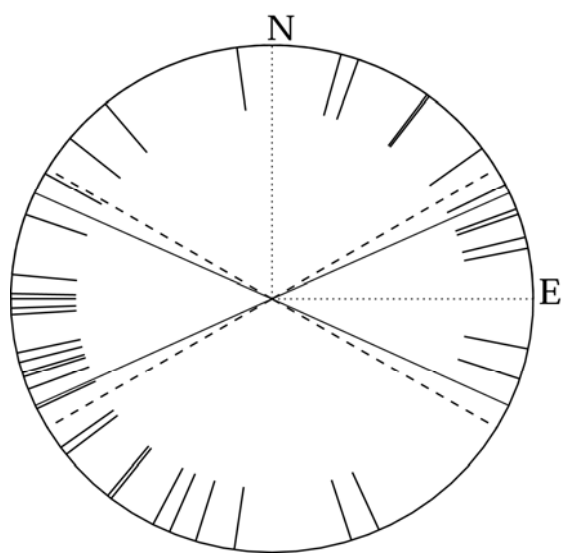

*Fig. 1: Orientation diagram for the heritage Andean churches and chapels of the Arica and Parinacota region, obtained from the data in Table 1. One can see there is a great diversity of orientations; despite this, a significant number (slightly less than half) follows a pattern of orientation within the solar range.*

**Orographic features**

In our analysis we took into account that in many of the sites we explored geographical conditions are unique and certainly have played a role when choosing the location of the churches; for example, those placed along rivers in some of the deep valleys of the region (see Gangui et al. 2016c for more details). Moreover, apart from these riverbeds and streams, numerous volcanoes and snowy peaks can serve as points of reference –like tutelary hills, even related to ancestor worship– when deciding the location and orientation of the temples, so we also checked our data to verify those possible orographic features.

The links between the local communities and the sacred spaces of the Andes, such as the *apus* (Reinhard 1983) or *achachilas* (which refers to "grandfather" or "ancestor"), are of great importance for the cultural and religious life of the inhabitants of the area (Leoni 2005; Martínez 1983). Also, recent ethnographic works, in spite of the known depopulation of the region, have revealed interesting connections between the communities and the cult of prominent hills (Choque and Pizarro 2013), elements that have survived in time linked to their cosmovision.

In the case of the chapel Virgen del Carmen, in Ungallire, as one emerges from the inside, on the very front, one finds the stunning view of the volcanoes Pomerape and Parinacota (the so called *payachatas*); these are located with azimuths just a few degrees on either side of the main axis of the construction. We verified something similar with the churches Virgen de la Inmaculada

Concepción, of the two villages Guallatire and Ancuta: according to our measurements, the Guallatiri volcano is nearly aligned with the axes of these churches but, in both cases, as it happened with the Ungallire chapel, the location of the volcano is in front of the buildings (Fig. 2).

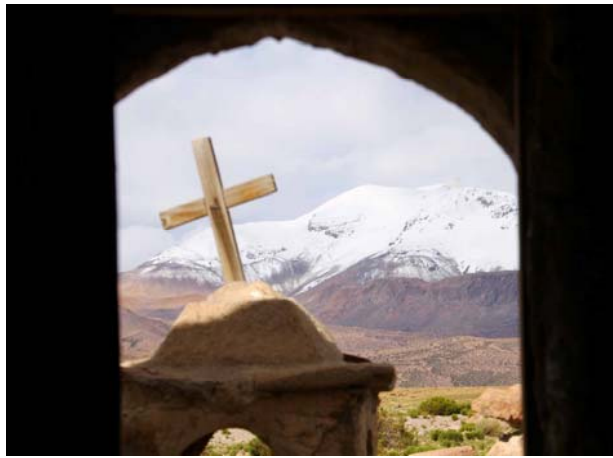

*Fig. 2: As one exits the chapel Virgen de la Inmaculada Concepción of Ancuta, one faces –with a high degree of precision– the Guallatiri volcano.*

If we now include the angular height of the horizon (h) for each sacred building in the analysis, we get the declination histogram (see Fig. 3). It shows two statistically –but not too notorious– significant peaks above the 3-sigma level (overall, there are three peaks within the solar range). Note that the frequency of events outside the lunisolar range (vertical lines) does not show significant peaks and suggests that a non negligible number of churches were constructed with orientations different from those dictated by the canonical rules (Gangui et al. 2016b).

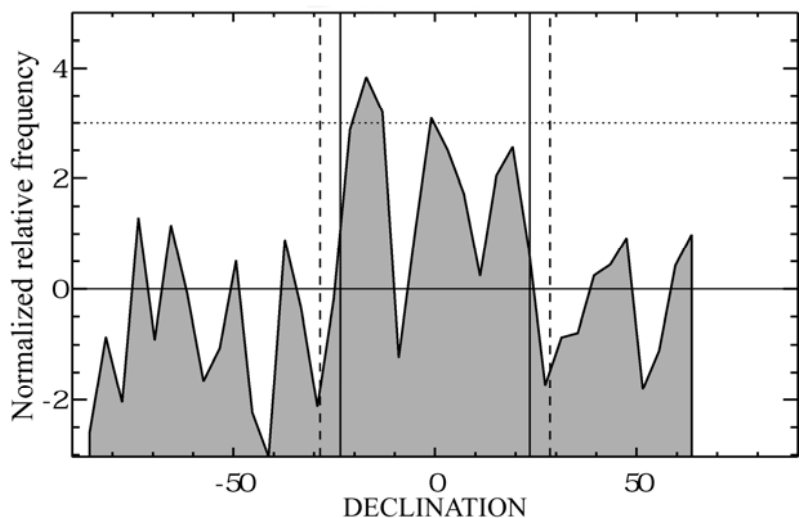

*Fig. 3: Declination histogram for the Andean churches of the Arica and Parinacota region. There are two statistically significant peaks, above the 3σ level, and not very noticeable (in total, there are three peaks within the solar range). Continuous vertical lines represent declinations corresponding to the extreme positions of the Sun at the solstices, while the dashed vertical lines represent the same for the Moon in major lunar standstills (lunistices). The frequency of declinations present outside the lunisolar range does not show significant peaks and suggests that a large number of constructions were oriented following non-canonical guidelines.*

## Conclusions

Our results show that, unlike what is commonly found in studies involving European historical churches, in the temples we studied here we found no single orientation pattern valid throughout the whole region. However, we found that almost half of the churches surveyed were oriented within the solar range, with a dominant share in those presenting their altar towards the west. We have also identified some notable cases where the orientation of the temples seems to follow the location of distinctive elements of the landscape –as volcanoes or other aboriginal culturally relevant *apus*– rather than the rising or setting Sun during meaningful dates for the particular dedication of the churches, a fact that brings to mind more the Aymara worship (Bouysse-Cassagne 1987) than the 16th century *Instructions on Ecclesiastical Building* written by the Italian Cardinal Carlo Borromeo (1985).

We are aware that the group of churches in this region –even if their number was adequate for a statistical study such as the one performed here– does not exhaust the work that should be addressed in the near future. The region of Arica and Parinacota is culturally closely related to areas of western Bolivia and southern Peru, not only in terms of geography, but especially in relation to the customs and religiosity of its inhabitants, both of today and of centuries past. We have recently carried out a first exploratory study of the orientations of some old colonial churches surrounding the city of Arequipa, Peru, along with others located on the left bank of the Colca Valley (Gutiérrez et al. 1986). A preliminary analysis of the measurements of these new churches in southern Peru indicates that, most likely, the patterns of orientation found in northern Chile may replicate around Arequipa. We have verified, for example, that several churches of the Colca Valley are oriented in the solar range and, in particular, one of them, Nuestra Señora de la Asunción de Chivay, besides being oriented to the east, has the property that, when leaving the temple through its main door, the visitor faces the Hualca Hualca volcano with an accuracy of only five degrees in azimuth with respect to the axis of the church.

In summary, we think the study of the precise spatial orientations of the heritage Andean churches we have begun here –and presented briefly in this note– can be extended to neighboring regions and, thus, may lead us to the complete characterization of these emblematic sacred constructions of the South Andean territory.

## Acknowledgements

This work was partially supported by CONICET and the University of Buenos Aires, Argentina, and by Fundación Altiplano, Chile.